# Misfit-dislocation hierarchy governs sliding of asymmetric non-CSL grain boundaries

Kunqing Ding[1,+], Yazhuo Liu[1,+], Yin Zhang[2], Lihua Wang[3], Xiaodong Han[3], Ting Zhu[1,*]

[1]Woodruff School of Mechanical Engineering, Georgia Institute of Technology, Atlanta, Georgia 30332, USA

[2]School of Mechanics and Engineering Science, Peking University, 100871, Beijing, China

[3]Institute of Microstructure and Property of Advanced Materials, Beijing Key Lab of Microstructure and Property of Advanced Materials, Beijing University of Technology, Beijing, 100124, China

## Abstract

Grain boundaries (GBs) strongly affect the mechanical response of polycrystalline materials, yet most atomistic studies have focused on coincidence site lattice (CSL) boundaries. Motivated by in situ atomic-resolution observations, we investigate step-free sliding along asymmetric non-CSL tilt GBs in face-centered cubic (FCC) metals using atomistic modeling. In these incommensurate GBs, a dense array of non-glissile primary misfit dislocations accommodates local interfacial mismatch, whereas a more widely spaced array of glissile secondary misfit dislocations accommodates residual mismatch. Uniform-sliding calculations reveal two distinct quantities: the minimum GB structural periodicity $\boldsymbol{\lambda}$, defined by the repeating arrangement of primary GB misfit dislocations, and the slip vector $\mathbf{b}$, corresponding to the minimum displacement-shift-complete translation that restores an equivalent GB structure. Nonuniform sliding proceeds through the glide of secondary GB misfit dislocations, which carry $\mathbf{b}$ and transform successive boundary segments between crystallographically equivalent translation states. The characteristic spacing between secondary misfit dislocations defines a longer periodicity $\boldsymbol{\Lambda}$. For the representative {331}/{111} GB, each secondary misfit dislocation glides through a thermally activated two-step kink-pair mechanism below the athermal stress, advancing by one structural period $\lambda$. These results establish a unified crystallographic and dislocation-based framework for understanding stress-driven sliding in structurally complex asymmetric GBs.

*Corresponding author: ting.zhu@me.gatech.edu

[+]These authors contributed equally.

## 1. Introduction

Grain boundaries (GBs) separate adjacent grains in polycrystalline materials and significantly influence yield strength, tensile ductility, creep resistance, and fracture toughness [1, 2]. In fine-grained and nanocrystalline metals, the substantial GB volume fraction can make GB-mediated plasticity the dominant deformation mode [3]. Although atomic-scale microscopy and modeling have significantly advanced the understanding of GB structure and behavior [4-11], most studies have focused on coincidence site lattice (CSL) GBs [12, 13], particularly symmetric tilt GBs [14], because of their relatively simple crystallography and computational tractability [15].

Recent experimental and theoretical studies have broadened the understanding of GB deformation beyond ideal symmetric boundaries to include structurally complex GBs. Descriptions based on dislocations and disconnections have established that interfacial line defects can accommodate both relative grain translation and shear-coupled boundary migration [7, 16-18]. Attention has also shifted toward asymmetric incommensurate non-CSL GBs [17-23]. For example, high-resolution transmission electron microscopy (HRTEM) has revealed the atomic structures of 90° ⟨110⟩ tilt GBs, while atomistic modeling has indicated inclination-dependent sliding resistance in {100}/{110} and {111}/{112} boundaries [24-26]. Subsequent in situ nanomechanical experiments demonstrated nearly frictionless sliding along asymmetric {100}/{110} GBs [26]. Despite these advances, the crystallographic origins and thermally activated kinetics of step-free sliding in asymmetric non-CSL GBs remain insufficiently understood.

Engineering polycrystals predominantly contain asymmetric non-CSL GBs. In face-centered cubic (FCC) materials, asymmetric non-CSL tilt GBs containing at least one low-index terminating plane, such as {111} or {100}, are frequently observed [9, 10, 27-30]. Under applied stress, these boundaries can undergo sliding mediated by step-free GB dislocations [9] rather than by the more commonly studied step-bearing disconnections. This distinction raises several fundamental questions. What determines the structural periodicity of these complex interfaces? What is the minimum structure-preserving relative translation between adjoining grains, represented by an effective displacement-shift-complete (DSC) translation [2]? How is this translation localized and propagated by GB dislocations during stress-driven sliding rather than occurring through coherent motion of the entire interface?

To address these questions, we use atomistic modeling of FCC Ni bicrystals to characterize stress-driven sliding in asymmetric non-CSL tilt GBs. The study is motivated by the in situ atomic-resolution experiments of Wang et al. [9], who directly observed GB dislocation motion during sliding-dominated deformation. We examine asymmetric ⟨110⟩ tilt GBs containing at least one low-index terminating plane, including {331}/{111}, {100}/{111} and {100}/{110}. In these incommensurate boundaries, the ratio of the surface-lattice repeat distances along the sliding direction is irrational, giving rise to interfacial mismatch across two characteristic length scales. A dense array of primary GB misfit dislocations accommodates the local mismatch, whereas a more widely spaced array of secondary GB misfit dislocations accommodates the residual mismatch relative to a near-CSL reference. Uniform-sliding calculations are used to determine the minimum structural periodicity **λ** and the structure-preserving slip vector **b**, which are generally distinct in complex GBs [4]. We find that primary GB misfit dislocations undergo only local rearrangements and are effectively non-glissile. By contrast, secondary GB misfit dislocations propagate across multiple structural periods and carry **b**, thereby mediating nonuniform GB sliding. In a relaxed, extended boundary, sparsely distributed secondary GB misfit dislocations recur with a longer characteristic spacing **Λ**. In the representative {331}/{111} GB, each secondary GB misfit dislocation advances through a localized two-step kink-pair mechanism below the athermal stress. These results demonstrate how a hierarchy of primary and secondary misfit dislocations governs the structure and sliding behavior of asymmetric non-CSL GBs.

## 2. Model and results

### 2.1 Simulation methods

All atomistic simulations were performed using LAMMPS [31], and atomic configurations were visualized using OVITO [32]. Asymmetric ⟨110⟩ tilt GBs were constructed in FCC Ni bicrystals by orienting two crystals with a common ⟨110⟩ tilt axis and joining them along the prescribed GB planes. For each bicrystal, relative translations between the grains were sampled, and the resulting structures were relaxed by conjugate gradient energy minimization. The lowest-energy interfacial configuration among the sampled translations was then selected for subsequent sliding calculations. Periodic boundary conditions (PBCs) were applied along the out-of-plane tilt-axis direction. The lateral, upper, and lower surfaces were traction-free during GB relaxation.

The Pt GBs observed experimentally by Wang et al. [8] provide the structural and mechanistic motivation for this study. However, the classical and machine-learned Pt potentials that we examined did not reproduce atomic GB structures  consistent with the experimental observations. The angular-dependent potential (ADP) for Ni developed by Howells et al. [33] reproduced the relevant experimental GB structural motifs and was therefore used as the interatomic potential for the FCC model system. The quantitative dislocation-core structures and energy barriers reported below are specific to this Ni potential. Nevertheless, the crystallographic relationships among structure-preserving GB translation states identified in the Ni simulations are expected to apply more broadly to FCC metals.

Relaxed finite-length GB models were used to determine the GB structural periodicity **λ.** Primary GB misfit dislocations were identified by extra half-plane terminations at the interface and the repeat distance of their arrangement defined **λ**. A separate bicrystal model was then constructed for the uniform-sliding calculations. Its dimension along the sliding direction was set to an integer multiple of **λ,** while its dimensions along the other two directions remained unchanged. PBCs were applied along both in-plane GB directions to eliminate free side surfaces and thereby suppress localized inelastic deformation at their intersections with the GB. Free surfaces were retained along the GB-normal direction during structural relaxation. Unless otherwise stated, all other calculations used the finite-length GB models.

Uniform GB sliding was simulated by applying successive shear displacement increments to the top and bottom boundary layers. At each increment, the upper grain was displaced relative to the lower grain in a direction parallel to the GB plane. The bicrystal was then relaxed by conjugate gradient energy minimization while the imposed displacements of the external boundary layers were maintained. The resulting sequence of relaxed configurations was used to identify crystallographically equivalent GB translation states and determine the slip vector **b,** which is the Burgers vector of a secondary GB misfit dislocation.

To investigate secondary GB misfit dislocations generated by applied shear in a finite-length GB model, a single secondary misfit dislocation was first introduced by imposing a relative displacement equal to its Burgers vector **b** between the two grains on one side of the intended dislocation core. Subsequent relaxation localized the imposed interfacial translation. The transition region separating the translated and untranslated interfacial states defined the dislocation core. A second secondary GB misfit dislocation was then introduced using the same procedure to reveal the resulting two-dislocation structure and the interaction between the dislocations.

The activation pathway for the glide of a secondary GB misfit dislocation was characterized using the stress-controlled nudged elastic band (σ-NEB) method [34]. A constant force in the GB sliding direction was applied to atoms in the upper five atomic layers, while atoms in the lower five layers were fixed, thereby maintaining the prescribed shear stress during relaxation. The initial and final configurations corresponded to adjacent local energy minima separated by one elementary propagation event of the dislocation. Intermediate NEB images were initially generated through linear interpolation between these configurations and subsequently relaxed using the QuickMin algorithm.

Two σ-NEB model geometries were used. In the quasi-two-dimensional (2D) model, the bicrystal thickness along the ⟨110⟩ tilt axis was approximately 1 nm. The small thickness constrained equivalent atomic columns along the tilt-axis direction to undergo nearly identical displacements, thereby enabling resolution of the elementary glide pathway and intermediate metastable states. To examine spatially localized, thermally activated processes, the bicrystal was extended along the tilt-axis direction to form a three-dimensional (3D) model. The 3D σ-NEB calculations allowed the secondary GB misfit dislocation line to move nonuniformly along its length, enabling localized kink-pair formation and migration.

### 2.2 Structure of {331}/{111} GB

We focused on an asymmetric non-CSL {331}/{111} GB in an FCC Ni bicrystal. Its structure closely resembles that of the FCC Pt GB observed by Wang et al. [9] using in situ nanomechanical testing combined with HRTEM. Figs. 1a-1c show the atomic structure of this asymmetric ⟨110⟩ tilt GB, which has misorientation angle of 21.8°. Grains G1 and G2 are aligned along the $[\bar{1}10]$ tilt axis. Grain G1 terminates at a high-index (331) surface, whereas grain G2 terminates at a flat close-packed (111) surface. The bicrystal measures 13 × 18 × 1 $nm^3$ and contains 21,812 atoms. For comparison, Figs. 1d-1f present HRTEM images of the analogous experimental GB [9]. The two grains are aligned along the $[\bar{1}10]$ zone axis, and each white spot represents a projected $[\bar{1}10]$ atomic column. Grain G1 exhibits a high-index (331) surface with single-atom-height steps separated by short (111) and $(1\bar{1}1)$ facets as marked in Fig. 1d, whereas grain G2 exhibits a flat close-packed (111) surface. This configuration constitutes an asymmetric non-CSL tilt GB and differs from the CSL boundaries commonly considered in atomistic simulations.

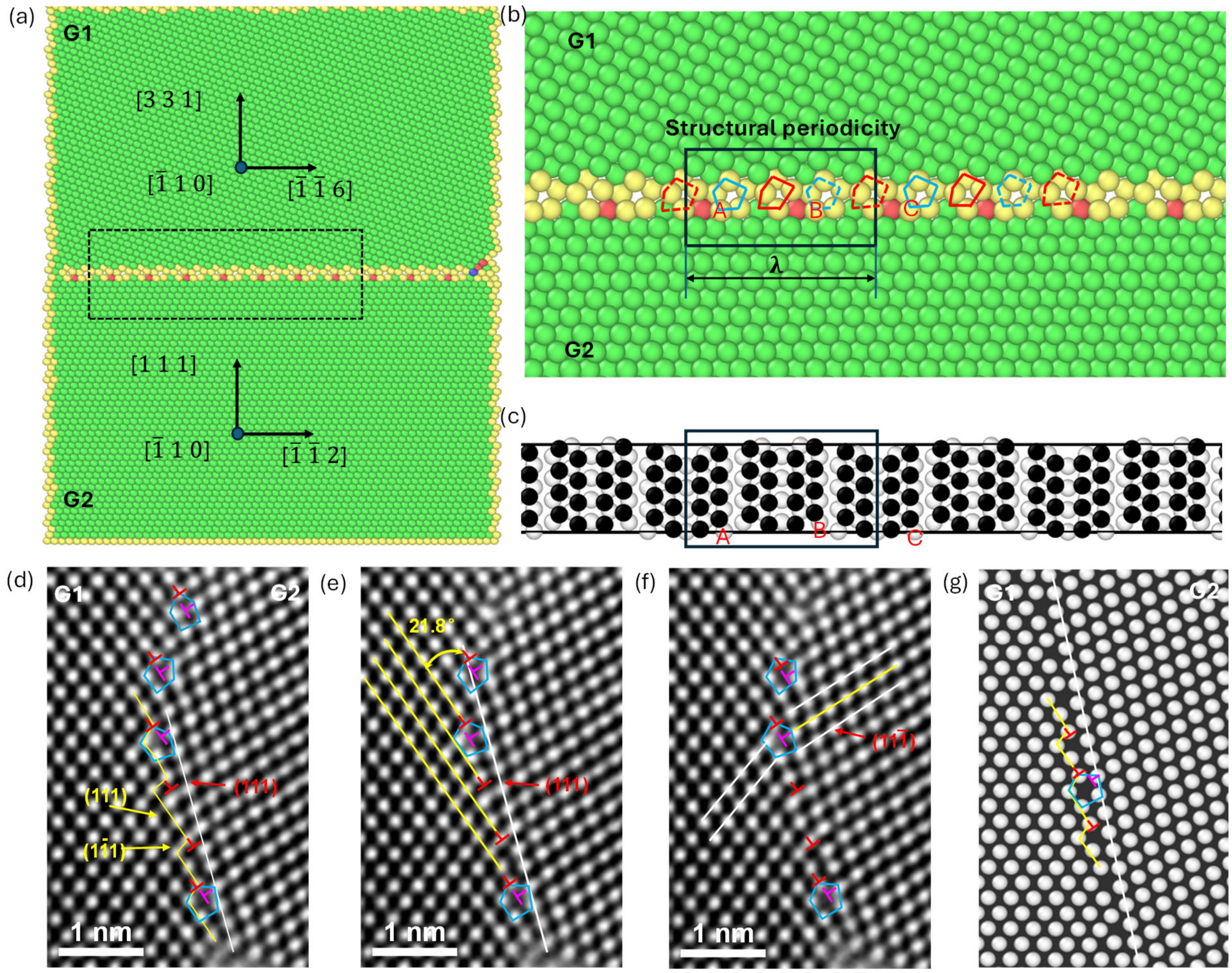

**Fig. 1.** Structure of an asymmetric non-CSL ⟨110⟩ tilt GB with a misorientation angle of 21.8°. (a) Atomic configuration of the Ni bicrystal. Grain G1 terminates at a (331) surface, and grain G2 terminates at a flat, close-packed (111) surface. Atoms are colored according to common neighbor analysis: FCC (green), hexagonal close-packed (HCP, red), and other local structures (yellow). (b) Enlarged view of the boxed region in (a). A structural period λ (solid box) spans four distinct pentagon-shaped structural units (outlined using different line colors and styles). (c) Top-down view showing only the grain surface atoms: G1 (black) and G2 (white). The out-of-plane registry of atoms A, B, and C (labeled in (a)) is shown along the $[\bar{1}10]$ direction. (d) HRTEM image of an analogous GB in a Pt bicrystal, adapted from Wang et al. [9]. G1 terminates at a high-index (331) surface with single-atom-height steps separated by short (111) and $(1\bar{1}1)$ facets (yellow lines), while G2 terminates at a flat (111) grain surface (white line). (e, f) The same HRTEM image as in (d), with yellow lines indicating extra half-planes associated with misorientation dislocations in (e) and a misfit dislocation in (f). Red ⊥ symbols identify the terminations of extra half-planes and thus the cores of misorientation dislocations. Pink ⊥ symbols identify the cores of misfit dislocations. (g) Visualization of the GB structure model using projected atomic columns.

To determine the GB structural periodicity, we analyzed the structural units within the boxed region of Fig. 1a (detailed in Fig. 1b). Although the projected boundary structure appears to contain alternating regular (blue) and distorted (red) pentagon units in Fig. 1b, these 2D projections do not define the complete periodicity because they omit the 3D interfacial registry. Fig. 1c presents a top-down view of the surface atoms in G1 (black) and G2 (white), revealing their relative registry along the $[\bar{1}10]$ tilt axis. The G1 atoms occupy threefold hollow sites above the close-packed (111) surface of G2. Atoms A and C have identical out-of-plane positions, whereas atom B is displaced to occupy a different out-of-plane position. As a result, the blue pentagon units associated with atoms A and B possess distinct 3D configurations, despite their similar projected profiles. The smallest complete structural repeat therefore contains four crystallographically distinct pentagon units (distinguished by color and line style in Fig. 1b). This repeat defines the structural periodicity $\boldsymbol{\lambda} = 5/2[\bar{1}\bar{1}2](111)$ relative to the G2 lattice, with a magnitude of about 2.2 nm.

The structural periodicity of an asymmetric non-CSL tilt GB is governed by the smallest repeating motif of primary GB misfit dislocations, including both their spacing and their sequence along the interface. In the experimental GB shown in Figs. 1d-1f, two types of interfacial features are interpreted as primary GB dislocations. The first accommodates lattice misorientation and thus is a primary GB misorientation dislocation with a Burgers vector of $1/2[110]$. It is marked by a red ⊥ symbol at the termination of an extra (111) half-plane indicated by a yellow line in Fig. 1e. The second accommodates interfacial misfit and thus is a primary GB misfit dislocation with a Burgers vector of $1/2[01\bar{1}]$. It is marked by a pink ⊥ symbol at the termination of an extra $(1\bar{1}1)$ half-plane indicated by a yellow line in Fig. 1f. These two dislocations frequently combine to form a GB Lomer lock. As shown in Fig. 1J of Ref. [9], Burgers circuit analysis identified a combined Burgers vector of $1/2[101]$ for the lock, consistent with $1/2[110] + 1/2[0\bar{1}1]$. In Figs. 1d-1f, the core of a GB Lomer lock appears as a regular pentagon, whereas the core of an isolated misorientation dislocation appears as a distorted pentagon. The experimental GB is not strictly periodic because it contains a step on the (331) surface. Identifying this step required careful examination of the boundary structure, given the small spacing between neighboring (311) planes [9]. In contrast, the model in Fig. 1g contains no such step and therefore isolates sliding along a flat boundary. This model exhibits a repeating sequence of alternating GB Lomer locks and misorientation dislocations. Its structural periodicity is determined by the spacing between successive crystallographically equivalent pentagons. This periodicity is consistent with the out-of-plane atomic registry discussed earlier.

The assignment of individual primary GB dislocations in the experimental HRTEM images should be interpreted cautiously. Figs. 1d–1f directly reveal atomic-column arrangements, extra half-planes, and variations in interfacial structural units. The primary GB misorientation and misfit dislocations, denoted by the red and pink ⊥ symbols, are assigned based on the terminations of extra half-planes and comparison with the corresponding atomistic model. This comparison is necessary because each white spot in the HRTEM image represents a projected atomic column, which can occupy one of two distinct positions along the out-of-plane $[\bar{1}10]$ direction, as discussed earlier. The experimental images alone therefore do not uniquely determine the 3D interfacial defect structure. The primary GB misorientation and misfit dislocations, together with the structural periodicity **λ** used in the subsequent analysis, are determined from the fully resolved atomistic model rather than from the experimental defect assignments alone.

### 2.3 Uniform GB sliding

Using the bicrystal model in Fig. 1a, we investigated uniform GB sliding. Within this GB, a dense array of primary GB misfit dislocations accommodated the local interfacial misfit, while the residual misfit was accommodated by elastic strain in the adjoining grains. Uniform sliding was imposed through incremental in-plane shear displacements of the upper boundary layer relative to the lower boundary layer. Conjugate gradient energy minimization was performed after each increment. This calculation provides a direct means to identify the structure-preserving translation and the corresponding GB slip vector **b**. Fig. 2a presents a magnified view of the GB region during uniform sliding and illustrates the leftward translation of grain G1 relative to G2. Structural analysis reveals that a full translation to a crystallographically equivalent site along the sliding direction occurs over four effective loading increments, each producing a discrete atomic displacement jump across the interface. Figs. 2b1-2b5 show top-down relative displacement maps for nearest-neighbor atom pairs across the GB at successive increments. The recurrence of the pattern from Fig. 2b1in Fig. 2b5 confirms completion of the four-increment sequence shown in Figs. 2b1-2b4. Within each map, neighboring atomic columns enclosed by the brown box exhibit identical relative displacements; four such column pairs constitute a single periodic unit, which is enclosed by the black box and is consistent with the structural periodicity identified in Fig. 1b. In each of Figs. 2b1-2b4, the average relative displacement over each periodic unit is approximately $1/8[\bar{1}\bar{1}2](111)$ relative to the G2 lattice. Fig. 2b6 shows the cumulative relative displacements of approximately $1/2[\bar{1}\bar{1}2](111)$ , identical for all nearest-neighbor atom pairs across the GB. Minor deviations from the exact lattice periods arise from elastic deformation caused by the applied shear load. Therefore, the structure-preserving DSC translation and the corresponding GB slip vector are $\mathbf{b} = 1/8[\bar{1}\bar{1}2](111)$.

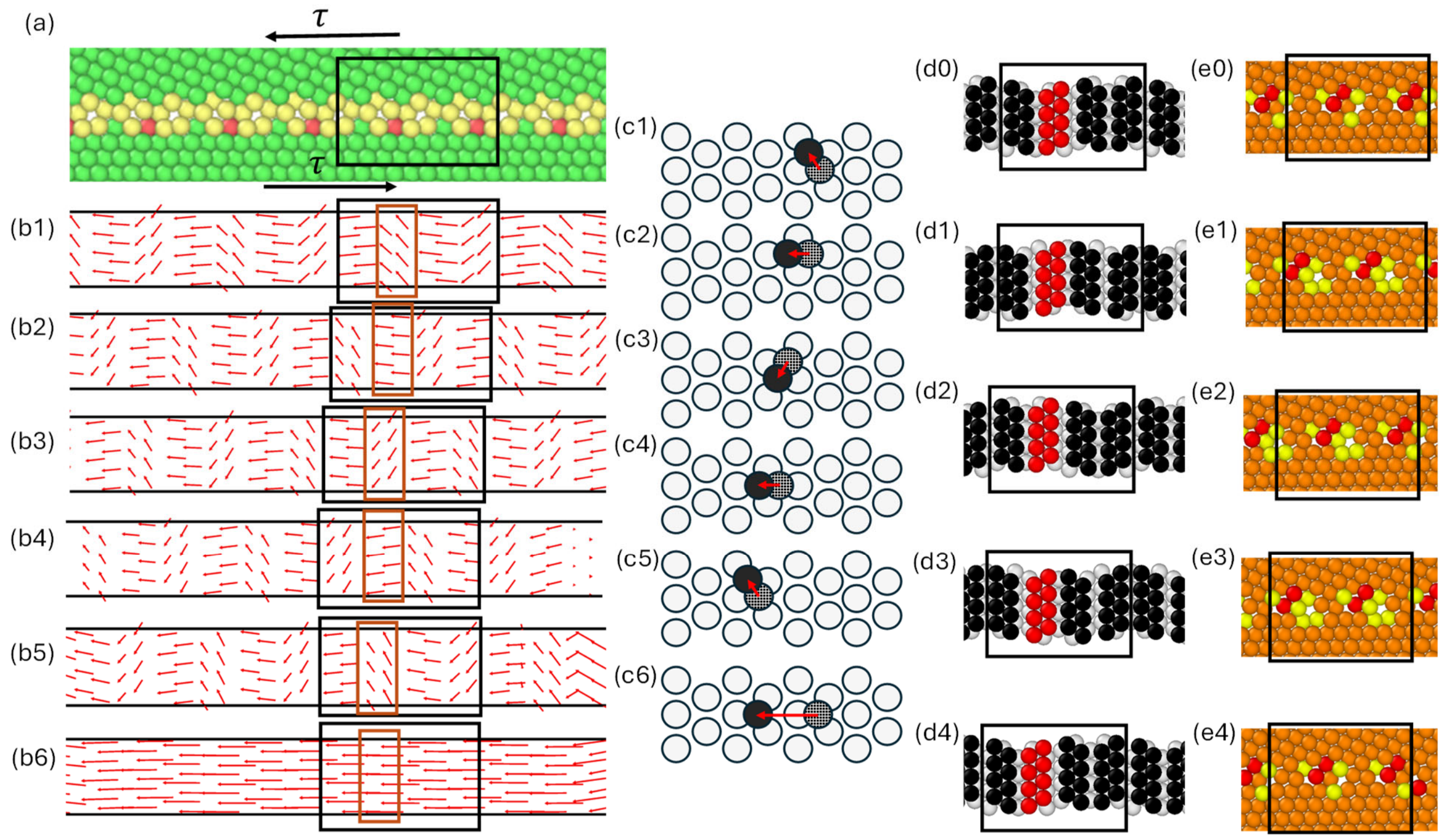

**Fig. 2.** Uniform GB sliding under an applied shear stress $\tau$. (a) Magnified view of the GB region, with one structural period indicated by a solid box. Atoms are colored according to common neighbor analysis. (b) Top-down relative displacement maps for nearest-neighbor atom pairs across the GB during four sequential sliding increments (b1-b4). The pattern in (b5) repeats that in (b1), confirming that a DSC translation is achieved after the four increments. Map (b6) shows the cumulative relative displacement map obtained by summing the displacements (b1-b4). (c) Schematics of a representative G1 surface atom progressing through four sequential sliding increments (c1-c4) to reach the DSC state (c5). The cumulative relative displacement of this atom is illustrated in (c6). (d) Top-down view of the positions of G1 and G2 surface atoms before and after each sliding increments (d0-d4). (e) $[\bar{1}10]$ views of the corresponding GB structures in (d). Atoms are colored by coordination number, $N$: 12 (orange, perfect FCC structure), 13 (red), and 11 (yellow).

In asymmetric tilt GBs containing a flat close-packed (111) grain surface, each surface atom of grain G1 undergoes four sequential increments to achieve a total translation of $1/2[\bar{1}\bar{1}2](111)$ along the sliding direction. Figs. 2c1-2c4 illustrate a G1 surface atom moving in $1/6\langle 112\rangle$ increments between threefold hollow sites (indicated by gridded to solid circles) atop the G2 (111) plane. After four increments, the atom reaches to a crystallographically equivalent site along the sliding direction (Fig. 2c5). The total relative displacement (Fig. 2c6) confirms a cumulative slip of $1/2[\bar{1}\bar{1}2](111)$. Each incremental displacement of an atomic column corresponds to a specific variant of the $1/6\langle 112\rangle$ family, and neighboring atomic columns within a single GB periodic unit follow different $1/6\langle 112\rangle$ variants. Nonetheless, the cumulative slip of $1/2[\bar{1}\bar{1}2](111)$ is determined by the G2 (111) lattice periodicity. Top-down views of the G1 and G2 surface atoms during these four sliding increments are shown in Figs. 2d0-2d4. In Fig. 2d4, the atomic pattern within the GB periodic unit (black box) repeats that in Fig. 2d0 but is translated leftward by $1/2[\bar{1}\bar{1}2](111)$. Figs. 2e0-2e4 present $[\bar{1}10]$ views for the same increments shown in Figs. 2d0-2d4, with atoms colored by coordination number. In Fig. 2e4, the periodic unit (black box) likewise repeats that in Fig. 2e0 with a shift of $1/2[\bar{1}\bar{1}2](111)$.

The uniform-sliding calculation characterizes the interfacial translation landscape and does not imply that an extended GB must slide coherently along its entire length. The initial and final configurations separated by DSC vector **b** are crystallographically equivalent translation states of the same GB. In an extended GB, this structure-preserving translation can be spatially localized. Adjacent GB segments may occupy the two translation states, and the line defect separating them constitutes a secondary GB misfit dislocation with Burgers vector **b**. As this defect propagates, it progressively transforms successive boundary segments from one translation state to the other. This process produces nonuniform GB sliding without requiring simultaneous translation of the entire interface. Such dislocation-mediated nonuniform sliding generally requires a lower driving stress than uniform sliding. The uniform-sliding analysis therefore provides a crystallographic basis for the subsequent study of nonuniform sliding mediated by GB dislocations.

### 2.4 Motion of secondary GB misfit dislocations

We next examined nonuniform GB sliding mediated by secondary GB misfit dislocations. In the asymmetric $\{331\}/\{111\}$ GB, a secondary GB misfit dislocation carries the Burgers vector **b** $=1/8[\bar{1}\bar{1}2](111)$. To produce the structure-preserving translation by **b**, the dislocation must propagate by at least one GB structural period $\lambda = 5/2[\bar{1}\bar{1}2](111)$. Thus, **b** defines the relative grain translation provided by dislocation glide, whereas $\lambda$ defines the propagation distance to restore an equivalent GB structure.

Primary and secondary GB misfit dislocations play distinct roles in the sliding process. Primary misfit dislocations constituent the interfacial defect structure and define its smallest repeating structural motif. During the sliding sequence in Fig. 2, they undergo local atomic rearrangements as the boundary transforms between equivalent translation states. Their cores do not propagate along the GB over distances exceeding one structural period $\lambda$ and are therefore effectively non-glissile within the GB plane. In contrast, a secondary misfit GB dislocation can glide successively across multiple structural periods. Its glide produces a relative grain translation **b** across the swept portion of the boundary, thereby generating a net sliding displacement. This distinction between locally rearranging primary misfit defects and propagating secondary misfit defects provides the dislocation-based foundation for understanding nonuniform GB sliding.

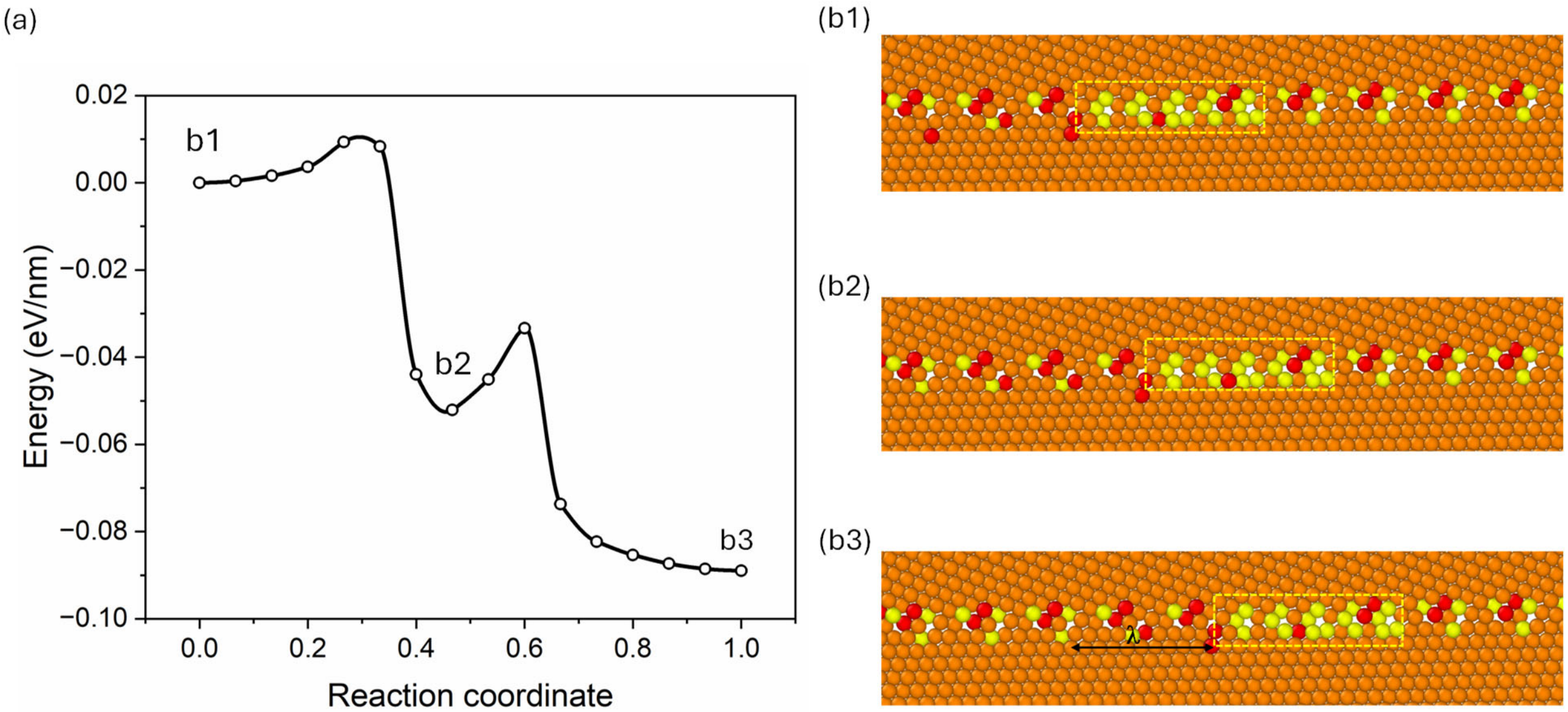

**Fig. 3.** Quasi-2D σ-NEB results for an elementary event of thermally activated glide of a secondary GB misfit dislocation. (a) MEP under an applied shear stress $\tau$ of 59.7 MPa. Three local energy minima, labeled b1-b3, indicate that a single elementary glide event proceeds through two sequential steps (b1→b2 and b2→b3). (b) Atomic configurations (b1-b3) correspond to these minima in (a). Atoms are colored by coordination number, as in Fig. 2e. The dislocation core, highlighted by a yellow dashed box, is composed primarily of miscoordinated (yellow) atoms and moves by one GB structural period $\lambda$ from (b1) to (b3).

We used the σ-NEB method [34] to investigate nonuniform GB sliding by the thermally activated glide of a single secondary GB misfit dislocation in a finite-length {331}/{111} GB. At finite temperatures and applied shear stress below the athermal threshold, dislocation glide requires thermal activation over an energy barrier. At the athermal threshold, this barrier vanishes and motion becomes instantaneous. To resolve the activation pathway, we first examined the GB dislocation in a thin bicrystal (about 1 nm thick). This geometry constrains the dislocation to quasi-2D motion, such that atoms within each $[\bar{1}10]$ column undergo nearly identical displacements (Figs. 3a-3b). To prepare for the σ-NEB calculation, a secondary GB misfit dislocation was introduced by imposing a relative displacement [35] of **b** between the two grains on the left side of the intended dislocation core, followed by structural relaxation. The resulting core (yellow box in Fig. 3b1) spans about five pentagon units and is composed primarily of miscoordinated atoms. The GB structure trailing the dislocation core differs from that ahead of the core, exhibiting variations in atomic coordination and out-of-plane registry because of the applied shear load. Configurations containing one and two secondary GB misfit dislocations are compared in the Discussion section.

At an applied shear stress $\tau$ of 59.7 MPa (below the athermal stress of 165 MPa), the σ-NEB calculation yielded a minimum energy path (MEP) for one rightward glide event of the secondary GB misfit dislocation. As shown in Fig. 3a, the MEP contains two successive energy barriers and three local minima, indicating that each propagation event proceeds through two sequential steps. The atomic configurations corresponding to these minima are shown in Figs. 3b1-3b3. Between the initial and final states, atom pairs within the dislocation core undergo a relative displacement equal to the Burgers vector **b** = $1/8[\bar{1}\bar{1}2](111)$. This shift results in a rightward translation of the dislocation core by one GB structural period $\lambda = 5/2[\bar{1}\bar{1}2](111)$ (Fig. 3b3). The intermediate local minimum (b2 in Fig. 3a) and its corresponding core configuration in Fig. 3b2 separate the two steps. The first step advances the core by $\lambda/2$, and the second step advanced it by the remaining $\lambda/2$. Repeated propagation of the secondary GB misfit dislocation through this stepwise glide mechanism produces nonuniform GB sliding. This result therefore identifies $\lambda/2$ (about 1.1 nm) as the smallest resolved core advances, whereas the complete event connecting equivalent endpoint states spans $\lambda$.

In the quasi-2D σ-NEB model, both energy barriers scale linearly with the dislocation-line length because the entire line is constrained to move coherently. This scaling cannot continue indefinitely as the line length increases, because a spatially localized 3D mechanism eventually becomes energetically favorable. The 3D σ-NEB calculation identifies such a mechanism, involving kink-pair formation and migration during the glide of a secondary GB misfit dislocation. This mechanism is uncommon in conventional FCC lattice-dislocation glide [34]. The bicrystal was extended to 17.9 nm along the out-of-plane $[\bar{1}10]$ direction, producing a model containing 392,778 atoms. Fig. 4a presents the corresponding MEP under an applied shear stress of 23.9 MPa. Figs. 4b1-4b4 show displacement contours for G1 surface atoms relative to the initial configuration. As in the quasi-2D σ-NEB calculation (Fig. 3), the 3D elementary glide event involves two sequential steps, each mediated by a kink pair. During the first step, a kink pair (Fig. 4b1) forms in the dislocation core near the first saddle point (b1 in Fig. 4a). The two kinks migrate in opposite directions until the dislocation regains a straight configuration at an intermediate minimum (Fig. 4b2; b2 in Fig. 4a). During the second step, another kink pair (Fig. 4b3) forms near the second saddle point (b3 in Fig. 4a). This pathway then reaches the final minimum (b4 in Fig. 4a), at which the dislocation core is again straight (Fig. 4b4). The barriers associated with saddle points b1 and b3 in Fig. 4a are 0.37 eV and 0.49 eV, respectively. Within the calculated pathway, the second barrier is larger and thus taken as the rate-limiting activation energy $E_{\mathrm{a}}$ for the elementary glide event. It may therefore control the kinetics of GB sliding when glide of a secondary misfit dislocation is the rate-limiting process.

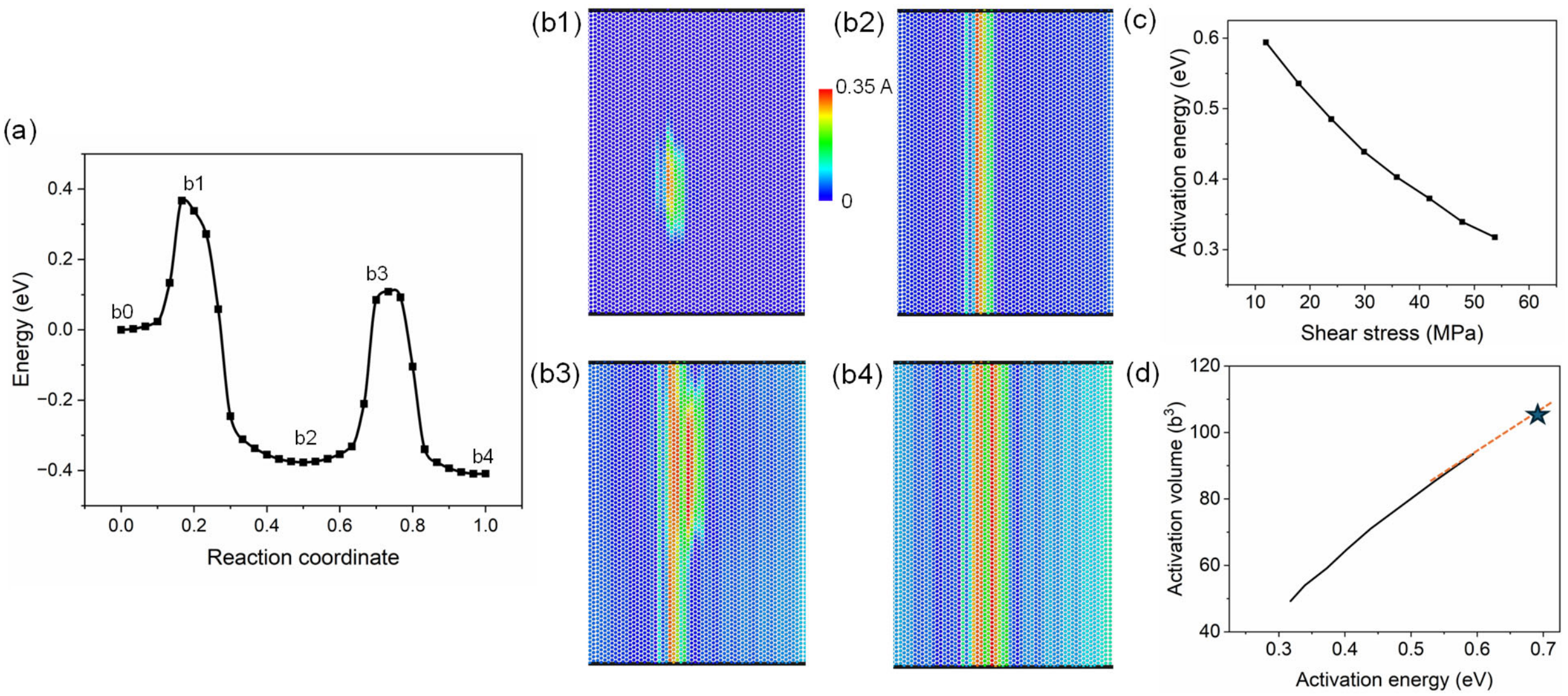

**Fig. 4.** 3D σ-NEB results for an elementary event of thermally activated glide of a secondary GB misfit dislocation. (a) MEP under an applied shear stress of $\tau = 23.9$ MPa. (b) Contour maps of the sliding displacement $\Delta d$ of G1 surface atoms, referenced to the unshown configuration at the initial local energy minimum (denoted b0). The corresponding energies of states b0-b4 are indicated in (a). (c) Activation energy $E_\mathrm{a}$ as a function of $\tau$. (d) Activation volume $V_\mathrm{a}$ as a function of $E_\mathrm{a}$. The data are extrapolated to higher $E_\mathrm{a}$ values (red dashed line) to estimate the characteristic $V_\mathrm{a}$ value (star) at $E_\mathrm{a} = 0.7$ eV.

Kink-pair formation is a well-established mechanism of thermally activated dislocation motion. For the representative $\{331\}/\{111\}$ GB, the present calculations show that kink-pair formation and migration involve a distinct rate-controlling atomistic pathway through which a localized GB translation state propagates across the periodic interfacial structure. The σ-NEB calculations characterize the associated energy barrier and reveal a mechanism comprising two sequential kink-pair steps. This mechanism is more complex than the single kink-pair step commonly associated with the elementary glide of an isolated lattice dislocation.

We further calculated the activation energy $E_\mathrm{a}$ for the kink-pair mechanism over a range of shear stresses $\tau$ (Fig. 4c). A characteristic $E_\mathrm{a}$ value of 0.7 eV corresponds to a typical laboratory strain rate of $10^{-3}$ $\mathrm{s}^{-1}$ at room temperature [34]. Extrapolation of the relationship in Fig. 4c suggests that the required $\tau$ is less than 10 MPa when $E_\mathrm{a} = 0.7$ eV. This result indicates that the secondary GB misfit dislocation may glide under relatively low driving stresses. Because of stability limitations, the 3D σ-NEB calculations did not converge at lower shear stresses corresponding to higher $E_\mathrm{a}$ values. We also determined the activation volume according to $V_\mathrm{a} = -\partial E_\mathrm{a}/\partial\tau$ [36]. Fig. 4d shows $V_\mathrm{a}$ as a function of $E_\mathrm{a}$ (black solid line), as derived from the $E_\mathrm{a}$ versus $\tau$ plot in Fig. 4c, with extrapolation to higher activation energies (indicated by the red dashed line). At $E_\mathrm{a} = 0.7$ eV, the extrapolated $V_\mathrm{a}$ is approximately $110b_\mathrm{L}^3$ (star in Fig. 4d), where $b_\mathrm{L}$ is the Burgers vector length of a lattice dislocation in FCC Ni. These atomistically determined activation volumes can be compared with experimental measurements [37] to evaluate strain rate sensitivity and identify potentially the rate-limiting process in polycrystals under typical laboratory strain rate conditions ($\sim 10^{-3}$ $\mathrm{s}^{-1}$) . However, direct comparison requires further model development that accounts for intragranular deformation processes and relates activation volumes measured in bulk polycrystals to those associated with localized GB-mediated events [38]. Finally, we note that although the sliding direction examined here is consistent with that used in the in situ experiments, sliding in the reverse direction exhibits a modestly different athermal stress. This directional asymmetry requires further investigation.

## 3. Discussion

### 3.1 Secondary GB misfit dislocations

To investigate the structure of coexisting secondary GB misfit dislocations, we sequentially introduced two such defects into a finite-length {331}/{111} GB containing no preexisting secondary misfit dislocations, as shown in Fig. 5a. First, we imposed a localized relative grain translation **b** [35] across the GB on the left side of the intended misfit dislocation core, followed by structural relaxation (Fig. 5b). We then increased the applied shear stress to drive this dislocation toward the right end of the bicrystal (Fig. 5c). Next, we imposed the same localized relative grain translation on the left side of the second dislocation core. The structure was subsequently relaxed while the shear load was maintained (Fig. 5d). Because the two dislocations have the same Burgers vector, their long-range elastic interaction is repulsive. A longer GB than that in Fig. 1a was required to accommodate both dislocations. The corresponding bicrystal measures 33 × 18 × 1 $nm^3$ and contains 55,252 atoms. This procedure can be extended to introduce multiple secondary misfit dislocations along a longer boundary. Alternatively, structural relaxation of an extended GB without applied shear can produce multiple coexisting secondary GB misfit dislocations, as shown in Section 3.2.

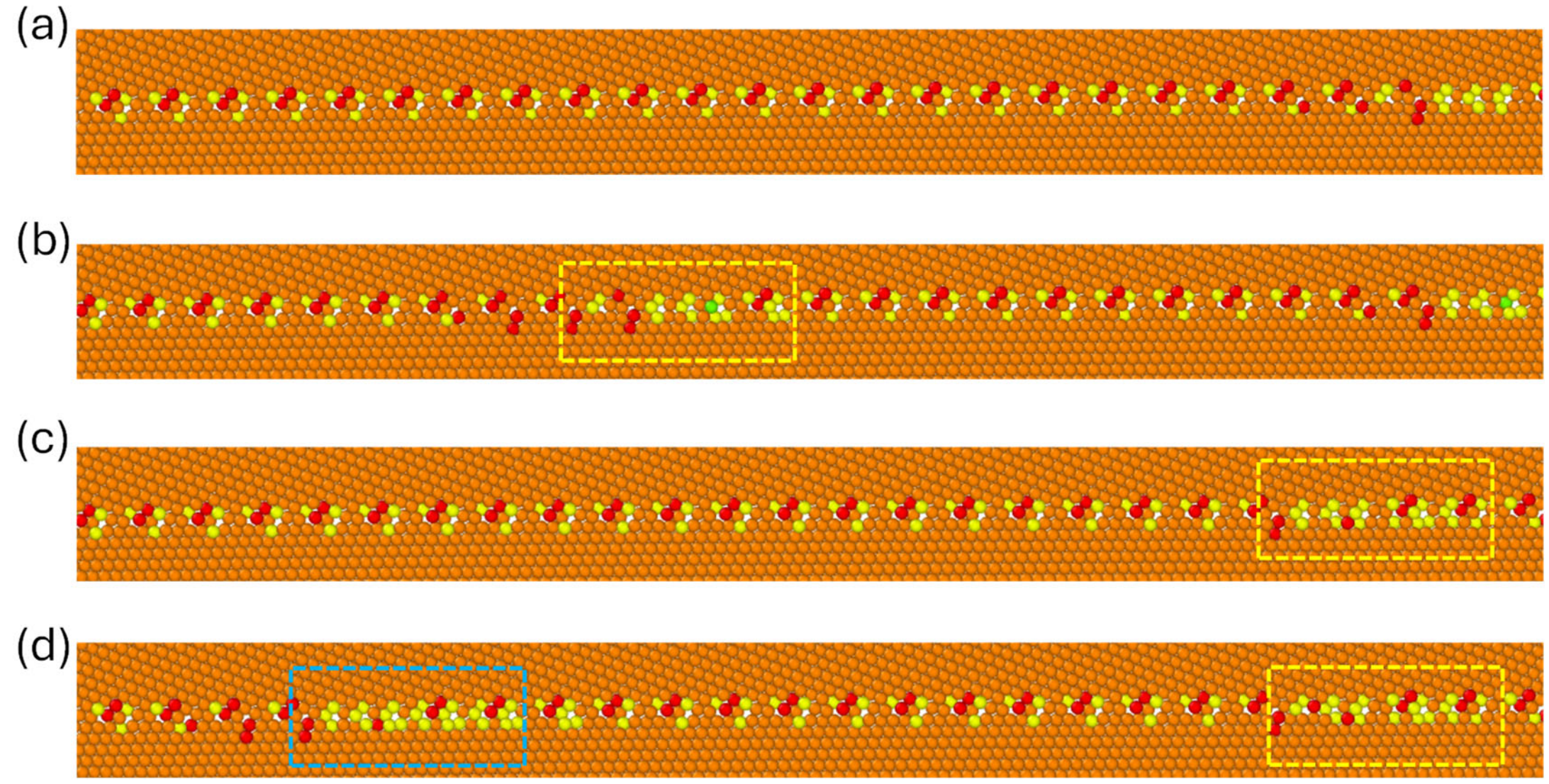

**Fig. 5.** Nonuniform sliding of an asymmetric ⟨110⟩ tilt GB with a 21.8° misorientation (the same structure as that in Fig. 1a) through the sequential glide of two secondary GB misfit dislocations. Atoms are colored by coordination number, as in Fig. 2e. (a) Initial GB structure prior to sliding, with no secondary GB misfit dislocations. (b) Introduction of the first dislocation carrying **b** under an applied shear stress. Its core is enclosed by a yellow box. (c) Rightward glide of the first dislocation under increasing shear stress. (d) Introduction of the second dislocation carrying **b** under a further increase in shear stress. It core is enclosed by a blue box.

For the representative {331}/{111} GB, the structure-preserving interfacial translation is mediated by the glide of secondary GB misfit dislocations. The corresponding Burgers vector is specific to the boundary bicrystallography and thus not universal among asymmetric non-CSL GBs. Other bicrystallographic configurations can support secondary GB misfit dislocations with different Burgers vectors. Nevertheless, interpreting a secondary GB misfit dislocation as a localized carrier of a structure-preserving interfacial translation remains generally applicable.

### 3.2 Near-CSL GB and misfit dislocations

The structural periodicity $\boldsymbol{\lambda}$ of the asymmetric non-CSL {331}/{111} GB suggests the existence of an underlying near-CSL structure. Conventional CSL GBs are defined by exact coincidence sites in a dichromatic pattern formed by two rigid grains with a prescribed misorientation angle [7]. Here, we extend this framework to asymmetric non-CSL tilt GBs using a near-CSL construction [39]. Near-CSL sites are identified by applying a positional tolerance of $5\%b_{\mathrm{L}}$, thereby incorporating small positional offsets that may be accommodated by local atomic relaxation during GB formation. As illustrated in Fig. 6a, the dichromatic pattern for the bicrystal in Fig. 1a contains a diamond-shaped near-CSL motif outlined in pink. Boundary planes aligned with a diagonal or an edge of this motif are compatible with periodic interfacial structures. Specifically, the vertical and horizontal diagonals correspond to mirror planes for symmetric GBs with the same misorientation, while the pink edge defines the asymmetric GB plane analyzed in Fig. 1a. Along each pink edge, the {331} surface of grain G1 comprises four single-atom-height steps separated by short {111} facets and interfaces with the close-packed {111} surface of grain G2. This geometry produces a repeating motif containing four pentagon units and thus defines the structural repeat vector $\boldsymbol{\lambda}$, consistent with Fig. 1b. The short yellow arrow indicates the minimum effective DSC vector **b** in the GB plane associated with the near-CSL reference. The near-CSL construction therefore identifies the short structural repeat $\boldsymbol{\lambda}$ and provides a geometrical estimate of the effective DSC translation **b** for this complex non-CSL tilt GB. The source code used to analyze the dichromatic pattern is available on GitHub (https://github.com/Yazhuo-Liu/DichromaticMap).

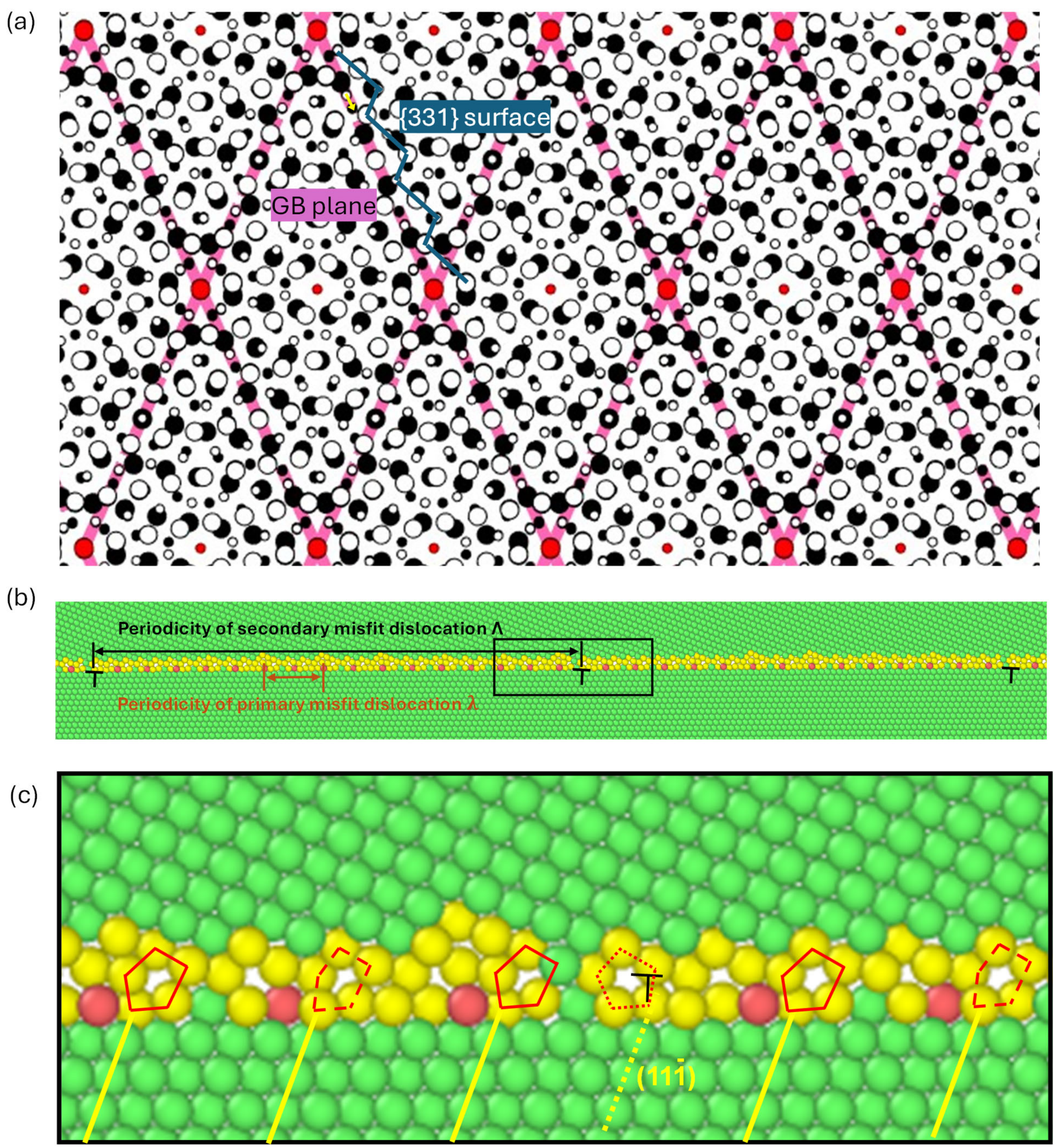

**Fig. 6.** Near-CSL analysis and long-range defect structure of the asymmetric {331}/{111} GB. (a) Dichromatic pattern of the bicrystal in Fig. 1a, revealing a diamond-shaped near-CSL motif with ∑40 and outlined in pink. GB planes aligned with the edges (thick pink lines) of the diamond are compatible with periodic interfacial structure. The pink edge defines the GB plane in Fig. 1a, along

which the {331} grain surface interfaces with the close-packed {111} grain surface. Within each repeat, the {331} surface contains four single-atom-height steps separated by narrow {111} facets (marked in blue). The short yellow arrow indicates the minimum effective DSC vector in the GB plane. (b) Atomic configuration of an extended GB. The short structural period $\lambda \approx 2.2$ nm corresponds to the repeating near-CSL structural unit. The black ⊥ symbols mark secondary misfit dislocations that accommodate the residual mismatch between the actual non-CSL boundary and the near-CSL reference. These dislocations have a mean spacing $\Lambda$ of 16.2 nm. The black box identifies the region enlarged in (c). (c) Magnified atomic structure of a secondary misfit dislocation in (b). Its core is outlined by a yellow dotted pentagon, and its associated extra half-plane is marked by a yellow dotted line. Neighboring primary misfit dislocation cores are outlined by solid and dashed pentagons, and their associated extra half-planes are marked by yellow solid lines. Panels (b) and (c) are colored according to common neighbor analysis.

The near-CSL description also clarifies the relationship between asymmetric non-CSL GBs and conventional CSL GBs. A symmetric CSL tilt GB can often be represented by a periodic array of primary GB misorientation dislocations. No primary misfit component arises because the terminating grain planes are related by symmetry. By contrast, an asymmetric CSL tilt GB can contain two primary GB dislocation sets. One set accommodates lattice misorientation, whereas the other accommodates the mismatch between the unequal terminating grain planes. The periodic arrangement of these dislocations defines the elementary boundary unit and its characteristic period. As the grain misorientation increases, the primary GB misorientation dislocations overlap and become difficult to resolve individually. However, primary GB misfit dislocations can remain identifiable from the disregistry between aligned close-packed atomic rows across the boundary. Because an exact CSL boundary is commensurate, no secondary GB misfit dislocation array is required.

The asymmetric non-CSL GBs considered here are incommensurate. A near-CSL reference comprising primary GB misfit dislocations with a repeat distance $\boldsymbol{\lambda}$ therefore cannot fully accommodate the lattice mismatch across the actual interface. The remaining mismatch is accommodated by secondary GB misfit dislocations, which can form in a relaxed extended boundary. Consequently, the boundary exhibits two characteristic length scales. The short period $\lambda$ is the repeat distance between neighboring equivalent primary GB misfit dislocations, which accommodate the local interfacial mismatch between adjoining grains. The long period $\Lambda$ is the characteristic spacing between secondary GB misfit dislocations, which accommodate the residual mismatch relative to the near-CSL reference. In the finite-length GB considered here, $\lambda$ and $\Lambda$ should be interpreted as characteristic spacings associated with a rational approximation to the incommensurate boundary rather than as exact periodicities of an infinite boundary. The value of $\Lambda$ is governed by the residual mismatch and may also depend on the system geometry and boundary conditions.

The two characteristic length scales are illustrated in Fig. 6b for a relaxed finite-length {331}/{111} GB. The red atoms indicate the cores of primary GB misfit dislocations, and the distance between every second red atom is $\lambda \approx 2.2$ nm. The secondary GB misfit dislocations are marked by black $\perp$ symbols and have a mean spacing of 16.2 nm, corresponding to $\Lambda$. Fig. 6c enlarges the boxed region in Fig. 6b and shows the atomistic structure of a secondary misfit dislocation. Its core is outlined by a dotted pentagon, and its associated extra half-plane is marked by a yellow dotted line. The neighboring primary GB misfit dislocation cores are outlined by solid and dashed pentagons, as in Fig. 1b, and their extra half-planes are marked by yellow solid lines. In regions containing only primary misfit dislocations, adjacent yellow solid lines are separated by four $(11\bar{1})$ atomic planes. Near the secondary misfit dislocation, the yellow dotted line is separated from each neighboring solid line by three $(11\bar{1})$ atomic planes. Hence, the dotted pentagon lies between two solid-outlined pentagons separated by six $(11\bar{1})$ atomic planes when the dotted extra half-plane is excluded from the count. This three-pentagon arrangement is consistent with the HRTEM observation in Fig. 4C of Ref. [9]. In addition, the local distribution of atomic coordination numbers around the secondary misfit dislocation differs from that around the neighboring primary misfit dislocations. Analysis of atomic disregistry shows that this defect has the same Burgers vector as the shear-induced secondary misfit dislocation examined in Fig. 5. The similarity of their core structures provides additional support for this correspondence. These observations indicate that the secondary misfit dislocations formed during relaxation of an extended GB have the same crystallographic defect character as the glissile secondary GB misfit dislocations induced by shear loading in Fig. 5. The present work focuses on stress-driven sliding involving repeating near-CSL structural units and a small number of secondary misfit dislocations in relatively short GBs. The collective interactions among multiple secondary misfit dislocations in extended GBs under shear loading remain for future study.

### 3.3 {100}/{111} GB

For asymmetric tilt GBs with larger misorientation angles, resolving individual GB misorientation dislocations becomes increasingly difficult because their cores begin to overlap. Nevertheless, the interfacial misfit component can still be identified from the atomic disregistry across the boundary. As an example, we examined a high-angle asymmetric $\langle 110 \rangle$ tilt GB previously characterized experimentally by Tian et al. [10]. It projected structure resembles that shown in Figs. 1d-1f, but its misorientation angle is larger at 54.7°. Based on the experimental structure, we constructed the corresponding Ni bicrystal model as shown in Fig. 7a. Both grains are aligned along the $[\bar{1}10]$ tilt axis. Grain G1 terminates at a low-index (001) surface, whereas grain G2 terminates at a close-packed (111) surface. The bicrystal measures 13 × 18 × 1 $nm^3$ and contains 22,079 atoms. As shown in Fig. 7b, the primary GB misfit dislocations can be identified from the interfacial disregistry. Their repeating arrangement defines the structural periodicity $\boldsymbol{\lambda} = 2[\bar{1}\bar{1}2](111)$ relative to the G2 lattice (Figs. 7b-7d). This periodicity differs from that of the {331}/{111} GB in Fig. 1b. By contrast, the larger misorientation prevents the misorientation component from being resolved as a discrete array of individual dislocations.

We next examined whether the structure-preserving slip vector changes when the misorientation and terminating plane of G1 are altered while G2 retains a flat (111) surface. Uniform-sliding calculations were performed using PBCs along the sliding direction. A full translation to a crystallographically equivalent site along the sliding direction occurs over seven effective loading increments, each producing a discrete load drop and yielding the slip vector **b** = $1/14[\bar{1}\bar{1}2](111)$ relative to the G2 lattice. Figs. 7c1-7c7 show top-down relative displacement maps for nearest-neighbor atom pairs across the GB during seven successive sliding increments. Fig. 7c8 shows the corresponding cumulative displacement of $1/2[\bar{1}\bar{1}2](111)$. During each increment, the average relative displacement over each structural period is approximately **b**, which corresponds to the Burgers vector of a secondary GB misfit dislocation. The configurations in Figs. 7d0-7d7 show that not all G1 surface atoms occupy ideal threefold hollow sites above the close-packed G2 surface. These results indicate that variations in misorientation and the opposing terminating plane can alter both $\boldsymbol{\lambda}$ and **b**, even when the close-packed {111} surface is retained. The characteristic spacing between secondary GB misfit dislocations defines the longer periodicity $\boldsymbol{\Lambda}$. We determined $\boldsymbol{\Lambda}$ by relaxing an extended {111}/{100} boundary without PBCs along the sliding direction. The resulting values of $\lambda$ and $\Lambda$ are approximately 1.7 nm and 6.5 nm, respectively.

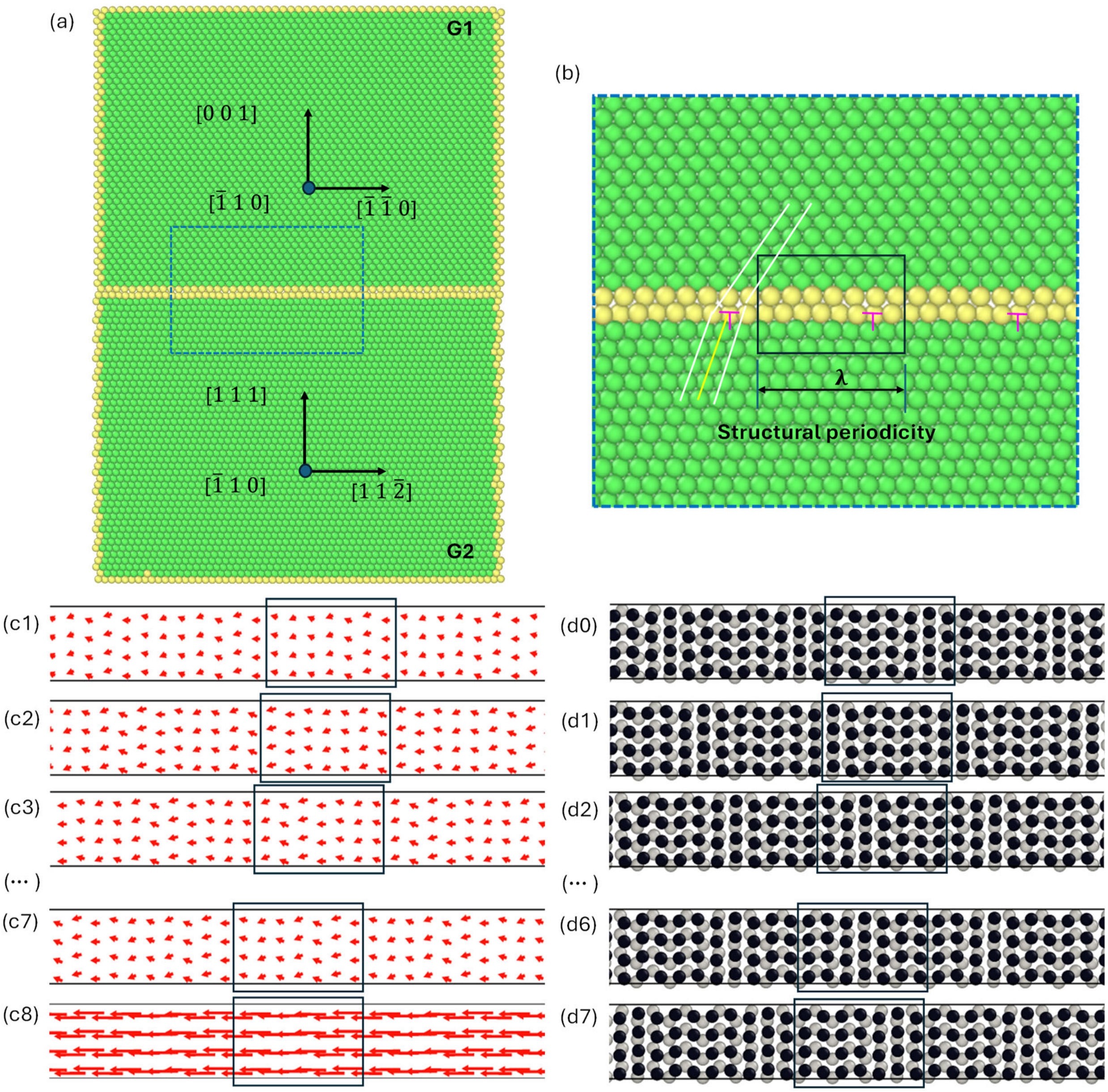

**Fig. 7.** {100}/{111} GB: an asymmetric ⟨110⟩ tilt GB with a 54.7° misorientation. (a) Atomic configuration of a Ni bicrystal: grain G1 terminates at a (001) surface, while grain G2 terminates at a close-packed (111) surface. (b) Enlarged view of the boxed region in (a). Primary misfit dislocations (marked by pink ⊥ symbols) are identified by the terminations of an extra half-plane, highlighted by the yellow line. The structural periodicity $\lambda$ is indicated by the black solid box. (c) Top-down relative displacement maps for nearest-neighbor atom pairs across the GB during seven sequential sliding increments (c1-c7). Map (c8) shows the cumulative relative displacement map (the sum of c1-c7). (d) Top-down views of the G1 and G2 grain surface atoms before and after

each of the seven sliding increments. More detailed top-down views of the relative displacement maps and atomic positions are provided in Supplementary Fig. S1.

### 3.4 {100}/{110} GB

The two GBs considered above both contain a $\{111\}$ terminating plane. We therefore examined whether the sliding characteristics $\boldsymbol{\lambda}$ and **b** depend on the presence of a close-packed $\{111\}$ plane. We constructed a crystallographically distinct asymmetric $\langle 110\rangle$ tilt GB with a 90° misorientation. As shown in Fig. 8a, grain G1 terminates at a $(\bar{1}\bar{1}0)$ surface, whereas grain G2 terminates at a (001) surface. The simulated GB structure is consistent with the direct HRTEM observations reported by Gautam et al. [25]. This configuration differs substantially from the $\{331\}/\{111\}$ and $\{100\}/\{111\}$ boundaries. Nevertheless, the $\{100\}/\{110\}$ GB exhibits well-defined structural periodicity, as shown in Fig. 8b. The terminations of extra half-planes reveal a set of primary GB misfit dislocations whose repeating arrangement defines the structural periodicity $\boldsymbol{\lambda} = 17/2[\bar{1}\bar{1}0](001)$ relative to the G2 lattice. This result further supports the conclusion that the GB structural periodicity is defined by the repeating arrangement of primary GB misfit dislocations. The detailed dislocation arrangement differs from those of the $\{331\}/\{111\}$ and $\{100\}/\{111\}$ boundaries because of their different bicrystallographies.

We next examined uniform sliding of the $\{100\}/\{110\}$ GB using PBCs along the sliding direction. A full translation to a crystallographically equivalent site along the sliding direction occurs over twelve effective loading increments, each producing a discrete load drop. Figs. 8c1–8c12 show top-down relative displacement maps for nearest-neighbor atom pairs across the GB during twelve successive sliding increments. Fig. 8c13 shows the cumulative displacement of $1/2[\bar{1}\bar{1}0](001)$. The corresponding atomic configurations before and after each increment are presented in Figs. 8d0–8d12. After the twelfth increment, the atomic arrangement within the marked periodic unit returns to a crystallographically equivalent configuration, indicating the completion of a full translation to the equivalent site along the sliding direction. In Figs. 8c1-8c12, the average relative displacement over a structural period is approximately $1/24[\bar{1}\bar{1}0](001)$, corresponding to the Burgers vector **b** of a secondary GB misfit dislocation in this boundary. The characteristic spacing between secondary GB misfit dislocations defines the longer periodicity $\boldsymbol{\Lambda}$. We determined $\boldsymbol{\Lambda}$ by relaxing an extended $\{100\}/\{110\}$ boundary without PBCs along the sliding direction. The resulting values of $\boldsymbol{\lambda}$ and $\boldsymbol{\Lambda}$ are approximately 4.2 nm and 14.4 nm, respectively.

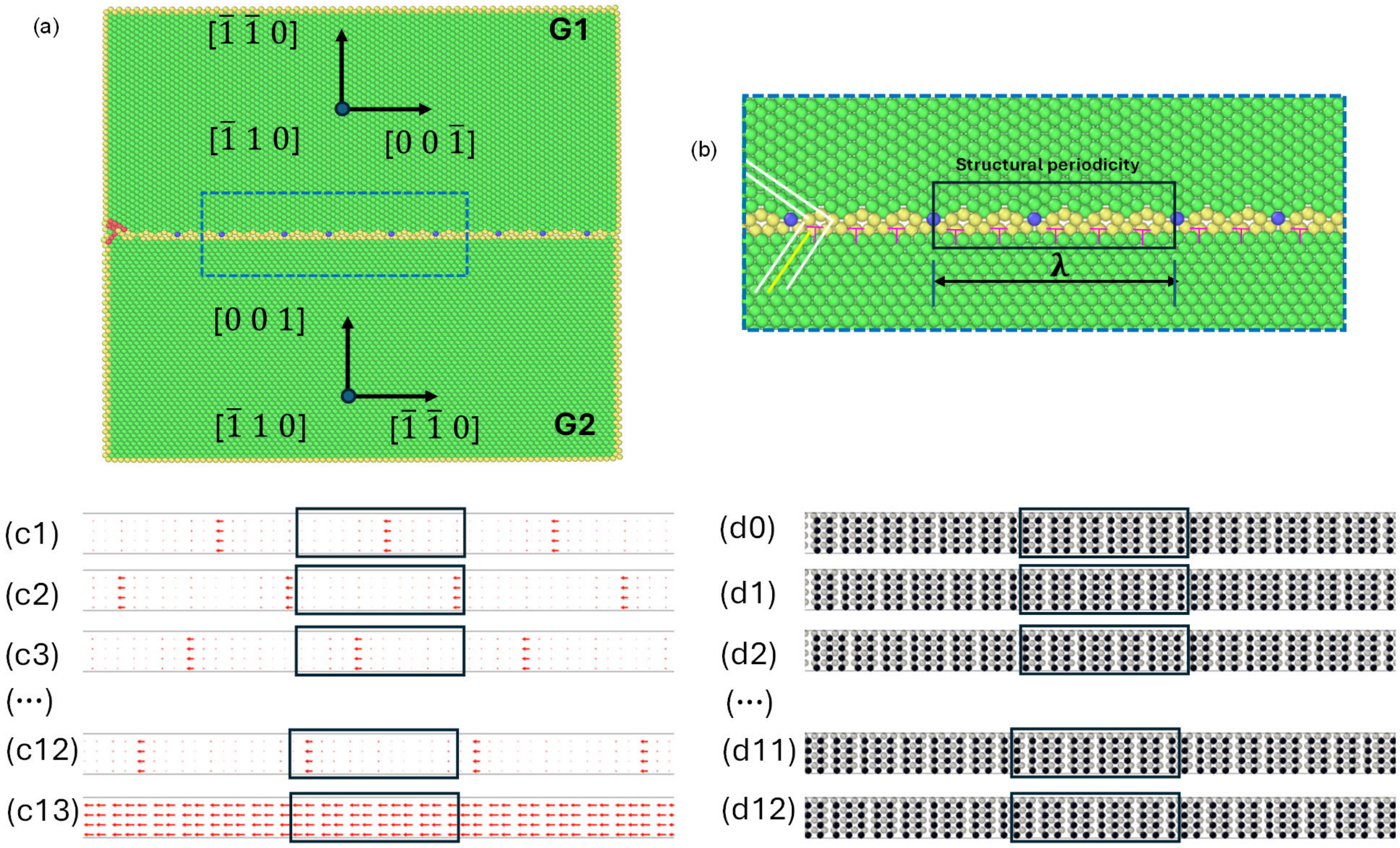

**Fig. 8**. {100}/{110} GB: an asymmetric ⟨110⟩ tilt GB with a 90° misorientation. (a) Atomic configuration of a Ni bicrystal: grain G1 terminates at a ($\bar{1}\bar{1}0$) surface, while grain G2 terminates at the more close-packed (001) surface. (b) Enlarged view of the dashed box in (a). Primary GB misfit dislocations (pink ⊥ symbols) are identified by the terminations of extra half-planes, highlighted by yellow lines. The structural periodicity **λ** is indicated by the solid box. (c) Top-down relative displacement maps for nearest-neighbor atom pairs across the GB during twelve sequential sliding increments (c1-c12). Map (c13) shows the cumulative relative displacement map (the sum of c1-c12). (d) Top-down views of the G1 and G2 grain surface atoms before and after each of the twelve sequential sliding increments. More detailed top-down views of the relative displacement maps and atomic positions are provided in Supplementary Fig. S2.

The sliding characteristics **λ** and **b** are therefore not universal among asymmetric non-CSL ⟨110⟩ tilt GBs. For the {111}-terminated GBs in Figs. 2 and 7, the respective slip vectors are **b** = 1/8⟨112⟩{111} and 1/14⟨112⟩{111}, while the respective structural repeat vectors are **λ** = 5/2⟨112⟩{111} and 2⟨112⟩{111}. By contrast, the {100}/{110} GB in Fig. 8 has **b** = 1/24⟨110⟩{100} and **λ** = 17/2⟨110⟩{100}. These comparisons demonstrate that **b** and **λ** are determined by bicrystallography.

These results also distinguish the bicrystallography-controlled sliding characteristics from material-dependent atomistic properties. The existence of structure-preserving interfacial translation states, the structural periodicity **λ**, and the Burgers vector **b** are determined by bicrystallography. By contrast, the core structures of secondary GB misfit dislocations, the critical stress, the activation energy and activation volume of kink pairs depend on the interatomic interactions. The quantitative kinetic results should therefore be regarded as predictions for the present Ni model rather than direct quantitative predictions for the Pt experiments [9]. The Pt experiments provide the structural and mechanistic motivation for studying GB-dislocation-mediated sliding, whereas the Ni calculations reproduce the relevant experimental GB structural motifs and resolve the underlying crystallographic and atomistic mechanisms.

**4. Concluding remarks**

We investigated stress-driven, step-free sliding in asymmetric non-CSL GBs in FCC Ni using energy minimization and NEB calculations. Motivated by atomic-resolution experimental observations, we examined asymmetric ⟨110⟩ tilt GBs containing at least one low-index terminating plane, including $\{331\}/\{111\}$, $\{100\}/\{111\}$ and $\{100\}/\{110\}$. The principal finding is that a hierarchy of misfit dislocations governs sliding along asymmetric non-CSL GBs. The supporting results are summarized below.

- Uniform-sliding calculations reveal discrete structure-preserving interfacial translation states and identify two characteristic quantities. The minimum GB structural periodicity $\boldsymbol{\lambda}$ is defined by the repeating arrangement of primary GB misfit dislocations. The slip vector **b** is the minimum DSC translation that restores an equivalent GB structure. Both quantities are determined by bicrystallography but characterize distinct aspects of the interface. These calculations establish a systematic procedure for determining $\boldsymbol{\lambda}$ and **b**.
- The extended asymmetric non-CSL GBs examined here are incommensurate [23] and contain two classes of misfit dislocations that accommodate distinct components of interfacial mismatch. Primary GB misfit dislocations accommodate the local interfacial mismatch between adjoining grains, whereas secondary GB misfit dislocations accommodate the residual mismatch relative to a near-CSL reference structure. The characteristic spacing between secondary GB misfit dislocations defines a longer periodicity $\boldsymbol{\Lambda}$, which can be determined by relaxing an extended boundary without PBCs along the sliding direction.
- Nonuniform sliding is mediated by secondary misfit dislocations. Primary misfit dislocations undergo local positional rearrangements but do not propagate over distances exceeding λ. They are therefore effectively non-glissile within the GB plane. In contrast, secondary misfit dislocations are glissile within the boundary and carry the structure-preserving translation vector **b**.

- For the representative {331}/{111} GB, the primary misorientation and primary misfit dislocations are lattice dislocations of the $1/2\langle 110\rangle\{111\}$ type. By contrast, a secondary misfit dislocation has the Burgers vector $\mathbf{b} = 1/8\langle 112\rangle\{111\}$ relative to the {111}-terminating grain. The short structural period λ is 2.2 nm, whereas the longer period Λ is 16.2 nm in the zeroth-order approximation [23]. Thermally activated propagation of a secondary GB misfit dislocation proceeds through localized kink-pair formation and migration. Each elementary propagation event advances the dislocation by λ through two sequential steps of λ/2. The structural repeat vectors and the Burgers vectors of secondary misfit dislocations were also determined for the {100}/{111} and {100}/{110} GBs, demonstrating that both quantities depend on boundary bicrystallography.

Several directions remain for future research. Although the present study focused on asymmetric tilt GBs containing at least one low-index terminating plane, general GBs can exhibit pronounced interface-energy anisotropy with respect to boundary inclination and can reduce interfacial energy by faceting into locally favorable orientations. Low-index and close-packed planes may therefore form constituent facets of general asymmetric GBs. The framework developed here can be extended to analyze sliding along such faceted interfaces. Furthermore, this study does not systematically compare sliding resistance among different GBs. Quantifying these resistances is necessary to determine how GB sliding competes with other intergranular and intragranular deformation mechanisms in polycrystals. In addition, the effects of shear loading on interactions between distributed secondary GB misfit dislocations warrant future study. More broadly, because GB sliding and interfacial friction both involve relative motion between adjoining lattices, the present results provide insight into friction between incommensurate crystals [23]. Overall, this work establishes a unified crystallographic and dislocation-based framework for understanding stress-driven sliding in structurally complex asymmetric GBs.

**Supplementary Information**

**Misfit-dislocation hierarchy governs sliding of asymmetric non-CSL grain boundaries**

Kunqing Ding[1,+], Yazhuo Liu[1,+], Yin Zhang[2], Lihua Wang[3], Xiaodong Han[3], Ting Zhu[1,*]

[1]Woodruff School of Mechanical Engineering, Georgia Institute of Technology, Atlanta, Georgia 30332, USA

[2]School of Mechanics and Engineering Science, Peking University, 100871, Beijing, China

[3]Institute of Microstructure and Property of Advanced Materials, Beijing Key Lab of Microstructure and Property of Advanced Materials, Beijing University of Technology, Beijing, 100124, China

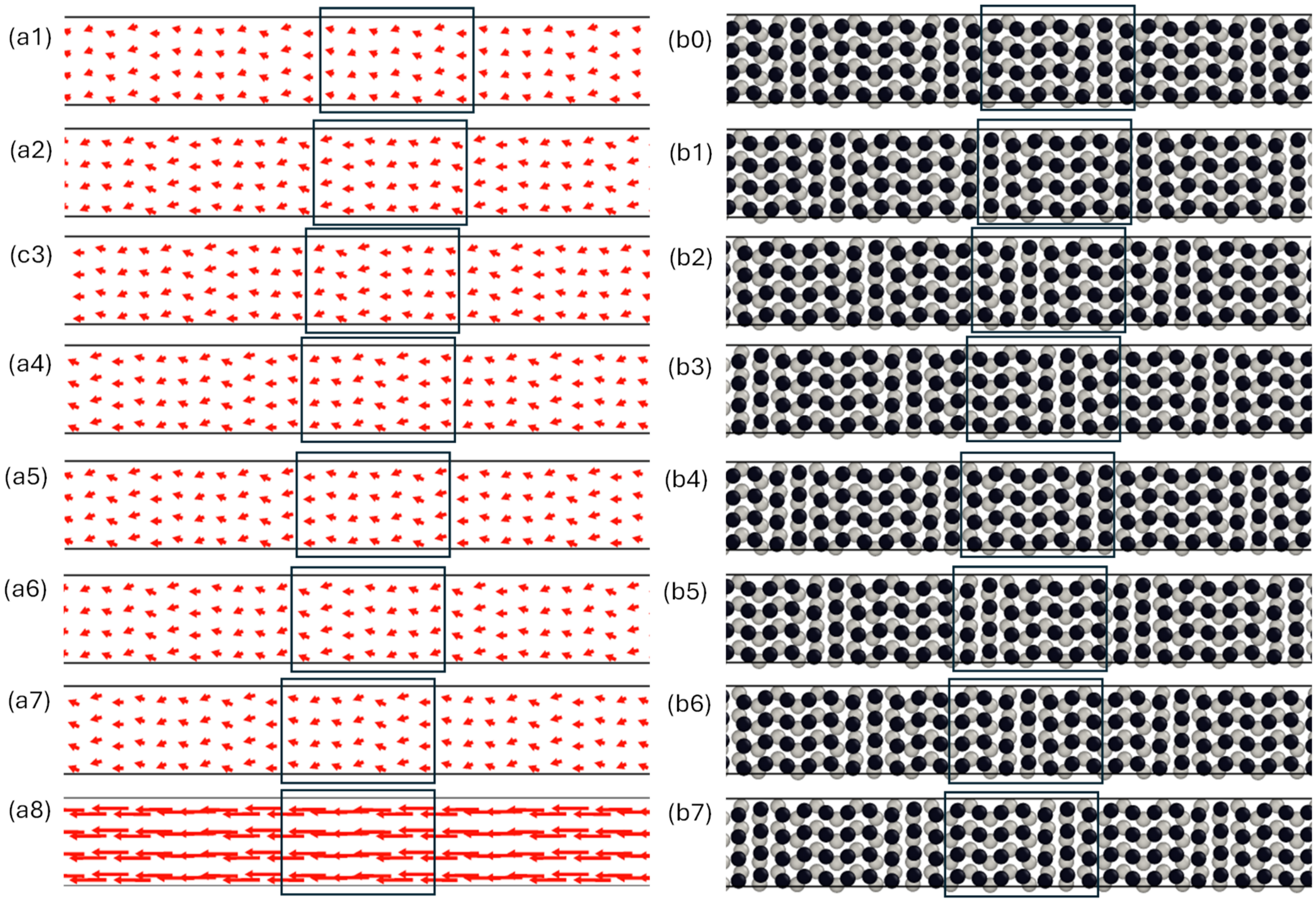

**Fig. S1.** {100}/{111} GB: an asymmetric ⟨110⟩ tilt GB with a 54.7° misorientation. (a) Complete top-down relative displacement maps for nearest-neighbor atom pairs across the GB during seven sequential sliding increments (a1-a7). Map (a8) shows the cumulative relative displacement map (the sum of a1-a7). (b) Complete top-down views of the G1 and G2 grain surface atoms before and after each of the seven sequential sliding increments.

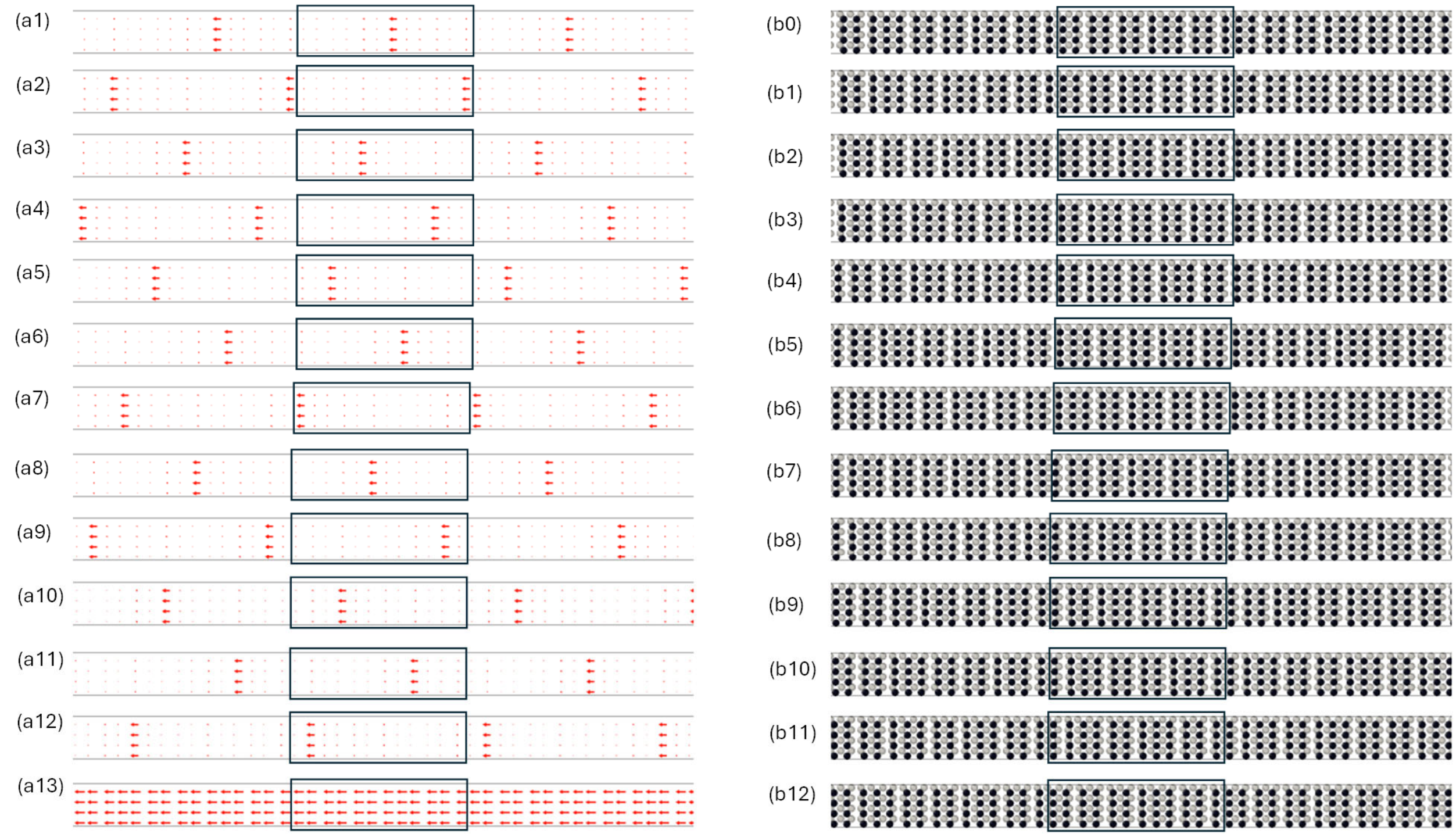

**Fig. S2.** {100}/{110} GB: an asymmetric ⟨110⟩ tilt GB with a 90° misorientation. (a) Complete top-down relative displacement maps for nearest-neighbor atom pairs across the GB during seven sequential sliding increments (a1-a12). Map (a13) shows the cumulative relative displacement map (the sum of a1-a12). (b) Complete top-down views of the G1 and G2 grain surface atoms before and after each of the seven sequential sliding increments.